\documentclass[aps,twocolumn,nofootinbib]{revtex4}
\usepackage[latin1]{inputenc}
\usepackage{epsfig}
\usepackage{color}

\begin{document}

\title{Boltzmann stochastic thermodynamics} 

\author{Mário J. de Oliveira}
\affiliation{Universidade de São Paulo,
Instituto de Física,
Rua do Matão, 1371, 05508-090
São Paulo, SP, Brasil}

\begin{abstract}

The Boltzmann kinetic equation is obtained from an integro-differential
master equation that describes a stochastic
dynamics in phase space of an isolated thermodynamic system.
The stochastic evolution yields a generation of entropy, leading
to an increase of Gibbs entropy, in contrast to a Hamiltonian
dynamics, described by the Liouville equation, 
for which the entropy is constant in time.  By considering
transition rates corresponding to collisions of two particles,
the Boltzmann equation is attained.
When the angle of the scattering produced by collisions is small,
the master equation is shown to be reduced to a differential equation
of the Fokker-Planck type. When the dynamics is of the Hamiltonian type,
the master equation reduces to the Liouville equation.
The present approach is understood as a stochastic interpretation
of the reasonings employed by Maxwell and Boltzmann in the kinetic
theory of gases regarding the microscopic time evolution.

\end{abstract}

\maketitle

\section{Introduction}

Stochastic thermodynamics
\cite{tome2006,zia2006,schmiedl2007,zia2007,seifert2008,blythe2008,
esposito2009,tome2010,broeck2010,tome2012,spinney2012,esposito2012,
zhang2012a,seifert2012,ge2012,santillan2013,luposchainsky2013,wu2014,
tome2015,tome2018},
or stochastic mechanics, extends Gibbs statistical mechanics to comprise
systems out of thermodynamic equilibrium. The extension is achieved
through the description of these systems by a stochastic dynamics,
meaning that
the time evolution in phase space consists of a stochastic trajectory 
rather than a pure deterministic trajectory. Usually the
stochastic dynamics is a continuous time Markovian process
and the time evolution of the probability distribution 
representing the whole system and defined 
over the phase space is governed by a master equation
or by a Fokker-Planck equation \cite{kampen1981,ebeling2005,tome2015book}.

A fundamental feature of stochastic thermodynamics is found in the
way of representing the production of entropy, 
which occurs in systems out of thermodynamic equilibrium.
The production of entropy is assumed to be related to the 
logarithm of the ratio of
the probability $P$ of a trajectory in phase space
and the probability of the reverse trajectory $P_r$.
The total entropy production $\Pi$ is obtained by
summing $P\ln(P/P_r)$ over all trajectories. 
The resulting expression is half of the summation of $(P-P_r)\ln(P/P_r)$
which is a nonnegative quantity, a fundamental property
of the entropy production, and a representation of the
second law of thermodynamics. In the nonequilibrium
situation, $P$ is distinct from $P_r$ making the
production of entropy a strictly positive quantity $\Pi>0$.
Thermodynamic equilibrium occurs when $P=P_r$
for all trajectories, implying the vanishing of the
entropy production, $\Pi=0$. The condition $P=P_r$
is the expression of microscopic reversibility or
detailed balance, and the hallmark of thermodynamic equilibrium.

For an isolated system, the time variation of entropy
equals the production of entropy implying, $dS/dt=\Pi$.
Since the production of entropy is nonnegative, 
it follows that the entropy increases monotonically
with time. This result has analogy with the Boltzmann
H-theorem according to which the quantity $H$ decreases
monotonically with time. In fact, the Boltzmann expression for
$-dH/dt$ is equivalent to the expression for $\Pi$.
The analogy extends to the Boltzmann kinetic equation
which is similar to a master equation,
revealing a connection between the kinetic theory of gases
\cite{maxwell1867,boltzmann1872,
watson1876,meyer1877,boltzmann1896,burbury1899,jeans1904}
and stochastic thermodynamics. 

The recognition that the reasonings employed by Maxwell and
Boltzmann within the kinetic theory of gases
were probabilistic and not strictly mechanical 
was clearly stated by Jeans \cite{jeans1904}.
However, their reasonings were not only probabilistic
in the static sense but also in the dynamical sense.
That is, their reasonings lead to the understanding that the time
evolution of the microscopic states of a
system was ruled by a stochastic dynamics.
This understanding was implicit in the views of
Ehrenfest concerning the irreversibility of the Boltzmann H-theorem,
when he presented an analogy between the 
Boltzmann kinetics and the dynamics of the urn model, which
is clearly a stochastic process \cite{ehrenfest1907}.

Here we set up a general master equation for the time
evolution of the probability distribution of the microscopic
states of systems of
interacting particles such as those studied within the kinetic
theory of gases. 
The allowed transitions are those corresponding to
the motion of two molecules, which are appropriate
to describe the collision of two molecules, employed
by Maxwell and by Boltzmann.
The Boltzmann kinetic equation 
\cite{huang1963,rumer1980,reichl1980,cercignani1988,kremer2010}
is then derived from the integro-differential master equation
by a truncation scheme in which a two-body
probability distribution is replaced by a product of
one-body probability distributions.

The transition rates describe the scattering occurring
in the interaction between two molecules. When the scattering
angle is small, the integro-differential Boltzmann equation
is shown to be reduced to a differential equation of the
Fokker-Planck type. Like the Boltzmann equation,
the differential equation preserves the kinetic energy.

In a stochastic dynamics, a trajectory starting from a
given point of the phase space is not unique. Each 
emerging trajectory has a certain probability of occurrence.
The multiple trajectories of the
stochastic evolution lead to the production of entropy. 
When the trajectory is unique, as happens to systems
described by a Liouville equation, on the other hand,
there is no production of entropy.

\section{Stochastic mechanics}

\subsection{General approach}

We consider a system of particles evolving in time
according to a dynamics defined in the phase space
$x=(q,p)$, where $q$ denotes the collection of coordinates $q_i$
of particles and $p$ the collection of momenta $p_i$.
During a certain time interval $\tau$, the state of the system
evolves from a state $x$ to a  state $x'$. 
If $x$ evolves to a single state $x'$,
as happens in a Hamiltonian dynamics, then
the motion is deterministic, that is, $x'$ is a function of $x$.
In a stochastic motion, $x$ can evolve to more than one state $x'$. 
To describe this situation, we resort to
a random variable $\xi$ by associating to each value of $\xi$
a certain value of $x'$, that is, $x'$ becomes a function of
$x$ and $\xi$, which we write as $x'=f(x,\xi)$.
A trajectory $x\to x'\to x''\to x'''\to\ldots$ in phase space 
becomes established if the sequence $\xi\to\xi'\to\xi''\to\ldots$ is given.
The random variables in this sequence are considered to be 
statistically independent, which means that the
stochastic process is Markovian.

If the system is in state $x$, we ask for the 
conditional probability ${\cal P}(x'|x)dx'$
of finding the system in phase space volume $dx'$
around $x'$ after an interval of time $\tau$. Denoting
by ${\cal P}(\xi)$ the probability density of $\xi$, 
the conditional probability is determined by
\begin{equation}
{\cal P}(x'|x) = \int \delta(x'-f(x,\xi)) {\cal P}(\xi)d\xi.
\label{19}
\end{equation}
The transition rate $w(x'|x)dx'$, defined as the rate in which the
state $x$ goes into state $x'$ around $dx'$, is proportional 
to the conditional probability, that is,
\begin{equation}
w(x'|x) = a(x) {\cal P}(x'|x),
\label{19a} 
\end{equation}
where the factor $a(x)$ might depend on the given state $x$.

Once the transition rate densities $w(x'|x)$ 
are given, we may write the master equation,
which is the equation that governs the time evolution of the
probability density $\rho(x,t)$ of $x$ at time $t$,
\begin{equation}
\frac{\partial}{\partial t}\rho(x)
= \int \{ w(x | x')\rho(x') - w(x' | x)\rho(x) \}dx',
\label{20}
\end{equation}
where we are omitting the dependence of $\rho$ on $t$.

In the stationary steady state, this equation becomes
\begin{equation}
\int \{ w(x | x')\rho^e(x') - w(x' | x)\rho^e(x) \}dx' = 0,
\label{20a}
\end{equation}
where $\rho^e(x)$ is the time independent stationary solution.
If, in addition, the transition rates $w(x'|x)$ obey the microscopic
reversibility, or detailed balance condition,
\begin{equation}
w(x | x')\rho^e(x') = w(x'|x)\rho^e(x),
\label{20b}
\end{equation}
for any pair of states $(x,x')$, the system is in
thermodynamic equilibrium and $\rho^e(x)$ is the equilibrium
probability density. If (\ref{20b}) does not hold, but (\ref{20a}) does,
the system will be found in a nonequilibrium stationary state 
and $\rho^e(x)$ is the nonequilibrium stationary probability density.

Let $E(x)$ denote the energy of the system when the state
is $x$. We are considering that $E(x)$ does not depend explicitly on time.
The time evolution of the average energy $U=\langle E(x)\rangle$,
defined by
\begin{equation}
U = \int E(x)\rho(x)dx,
\end{equation}
is obtained by multiplying equation (\ref{20}) by
$E(x)$ and integrating over $x$. The result is 
\begin{equation}
\frac{dU}{dt} = - \Phi_u,
\label{21a}
\end{equation}
where
\begin{equation}
\Phi_u = \int \{ E(x') - E(x)\} w(x|x')\rho(x')dx' dx,
\label{21b}
\end{equation}
is understood as the flux of energy from the system to the environment.
In general, the flux of energy is a sum of the heat flux and the 
work done on the system per unit time. In the absence of the last
term, $\Phi_u$ is just the heat flux which is understood as describing
the heat exchange by the contact with the environment and also by
absorption and emission of radiation. 
In deriving expression (\ref{21b}), we have rearranged the second term
by exchanging the variables $x$ and $x'$.

The entropy $S$ of the system is defined as the Gibbs entropy
\begin{equation}
S(t) = - k_{\mbox{\tiny B}} \int \rho(x,t) \ln \rho(x,t) dx,
\label{18}
\end{equation}
where $k_{\mbox{\tiny B}}$ is the Boltzmann constant.
Its time derivative can be written as \cite{tome2015}
\begin{equation}
\frac{dS}{dt} = \Pi - \Phi,
\label{22}
\end{equation}
where $\Phi$ is the flux of entropy from the system to the 
environment, 
\begin{equation}
\Phi = - k_{\mbox{\tiny B}} \int  w(x|x')\rho(x') \ln \frac{w(x'|x)}{w(x|x')} dxdx',
\label{22a}
\end{equation}
and $\Pi$ is the entropy being generated inside the system
per unit time, or the entropy production per unit time,
given by
\begin{equation}
\Pi = - k_{\mbox{\tiny B}}
\int  w(x|x')\rho(x') \ln \frac{w(x'|x)\rho(x)}{w(x|x')\rho(x')} dxdx'.
\label{22b}
\end{equation}
It is straightforward to show that $\Pi$ is a nonnegative quantity,
$\Pi\ge 0$. To this end, we exchange variable $x$ and $x'$ 
in the right-hand side of equation (\ref{22b}) to reach
an expression which, added to (\ref{22b}), gives us the equation 
\[
\Pi =  \frac{k_{\mbox{\tiny B}}}2
\int \{ w(x'|x)\rho(x) - w(x|x')\rho(x') \} \times 
\]
\begin{equation}
\times \ln \frac{w(x'|x)\rho(x)}{w(x|x')\rho(x')} dxdx'.
\label{23}
\end{equation}
The integrand is of the form $(a-b)\ln(a/b)\geq 0$, which is
nonnegative.

\subsection{Time reversal}

Suppose a system makes the transition
\begin{equation}
x=(q,p) \to x'=(q',p')
\end{equation}
during a certain interval of time $\tau$. The {\it time reversal} transition is
\begin{equation}
x_{\rm tr}'=(q',-p') \to x_{\rm tr}=(q,-p),
\end{equation}
which might be distinct from
\begin{equation}
x'=(q',p') \to x=(q,p),
\end{equation}
which we call the {\it reverse} transition.
In some cases, the reverse transition may not exist in which  case
the detailed balance condition (\ref{20b}) cannot be fulfilled.
But if, in this case, the time reversal transition exists, then
the detailed balance condition can be fulfilled in
the following form,
\begin{equation}
w(x_{\rm tr}| x_{\rm tr}')\rho^e(x_{\rm tr}') = w(x'|x)\rho^e(x),
\end{equation}
and again the system will be found in thermodynamic 
equilibrium. Nevertheless,
the formalism developed above can still be used if
we formally replace the time reversal transition by the
reverse transition and assume that the density $\rho(q,p)$
is even in the variable $p$.

\subsection{Isolated system}

An isolated system does not exchange heat and performs no
work so that its energy $E(x)$ remains invariant in time.
The conservation of energy in a transition $x\to x'$
is fulfilled by assuming that $w(x'|x)$ is nonzero
only when $E(x')=E(x)$.
This implies that the flux of energy, given by equation
(\ref{21b}), vanishes identically, $\Phi_u=0$.
Also, in an isolated system there should be no entropy flux. 
To fulfill this condition, we assume that
the rate of a transition is equal to the reverse
transition, that is, $w(x'|x)=w(x|x')$,
and the flux of entropy, given by 
equation (\ref{22a}), vanishes identically, $\Phi=0$.
The master equation (\ref{20}) in the absence of entropy flux becomes 
\begin{equation}
\frac{\partial}{\partial t}\rho(x) = \int w(x | x') \{\rho(x') - \rho(x) \}dx',
\label{20i}
\end{equation}
and, in addition, from equation (\ref{23}), we find
\begin{equation}
\Pi =  \frac{k_{\mbox{\tiny B}}}2 \int w(x'|x)\{\rho(x) - \rho(x') \}
\ln \frac{\rho(x)}{\rho(x')} dxdx'.
\label{51}
\end{equation}

The vanishing of the two quantities $\Phi$ and $\Phi_u$ characterizes
an isolated system.
In this case, the detailed balance condition (\ref{20b}) gives
$\rho^e(x)=\rho^e(x')$ if $E(x)=E(x')$ which implies that
the equilibrium probability distribution is of the form 
\begin{equation}
\rho^e(x) = F(E(x)),
\label{25}
\end{equation}
that is, $\rho^e(x)$ depends on $x$ through $E(x)$. 

Let us consider as the initial condition
a probability distribution such that all states $x$
have the same energy $E_0$, that is, 
\begin{equation}
\rho(x,0)\neq 0 \qquad {\rm only\,\,\,if} \qquad E(x)=E_0.
\label{27}
\end{equation}
For example, $\rho(x,0)=\delta(x-x_0)$ where $E(x_0)=E_0$.
Since the energy is conserved by the
transition rates, it follows that any subsequent probability
distribution will hold the same property (\ref{27}), including
the stationary probability distribution.
We may conclude from property (\ref{25}) that $\rho^e(x)=F(E_0)$,
that is, $\rho^e(x)$ has the same value for all states 
such that $E(x)=E_0$, or
\begin{equation}
\rho^e(x)=\frac1\Omega \delta(E(x)-E_0),
\label{26}
\end{equation}
which is the equilibrium Gibbs microcanonical probability distribution,
where
\begin{equation}
\Omega = \int \delta(E(x)-E_0)dx.
\end{equation}
The trajectories lie on the surface of constant energy
in phase space. However, the states will
be equiprobable and given by (\ref{26}), only for long times.

We remark that in the present case of an isolated system,
$dS/dt=\Pi\geq0$, since $\Phi=0$,
where $\Pi$ is given by (\ref{51}).
For any initial probability
distribution distinct from the final distribution $S$ will 
increase with time, as expected.

\subsection{Boltzmann equation and H-theorem}

We analyze the case in which 
the allowed transitions from a state $x$ to a state $x'$ are
those in which only two components of $x$, say $x_i$ and $x_j$,
have their states modified. That is, 
$(x_i,x_j)$ changes to $(x'_i,x'_j)$ whereas $x_k$, $k\neq i$
and $k\neq j$, remain unchanged. 
The corresponding transition rate is denoted by $w_{ij}(x'|x)$.
To reach the Boltzmann kinetic equation, 
we assume two conditions. One of them is
that the transition rate is equal to its reverse
$w_{ij}(x'|x)=w_{ij}(x|x')$. This describes the evolution
of an isolated system, as we have seen above.
The other condition is that the transition rate $w_{ij}(x'|x)$
depends only on the variables $i$ and $j$, and write
$w_{ij}(x_{ij}'\,|\,x_{ij})$ where the notation
$x_{ij}= (x_i,x_j)$ is being used.
This second condition is necessary to describe 
a binary collision, in which the states of two particles are being changed.

The transition rate $w(x'|x)$ in equation (\ref{20i}) becomes a
sum of all pairs $i,j$ of $w_{ij}(x_{ij}|x_{ij}')/n$,
\begin{equation}
\frac{\partial\rho(x)}{\partial t} = \frac1{n} \sum_{(ij)}\int w_{ij}(x_{ij} | x_{ij}') 
\{\rho(x^{ij}) - \rho(x) \}dx_i'dx_j',
\label{20j}
\end{equation}
where $n$ is the number of components of 
the vector $x$ and the summation extends over all distinct pairs $(ij)$.
The notation $x^{ij}$ stands for the vector with the same components
of the the vector $x$ except the $i$ and $j$ components which are
$x_i'$ and $x_j'$, respectively

Integrating the left and right-hand sides of equation (\ref{20j})
in all variables except $x_1$, we have
\[
\frac{\partial}{\partial t}\rho_1(x_1)
= \int w_{12}(x_{12} | x'_{12}) \times 
\] 
\begin{equation}
\times\{\rho_2(x'_{12}) - \rho_2(x_{12}) \} dx'_1 dx'_2 dx_2,
\label{46}
\end{equation}
where $\rho_1(x_1)$ is a one-variable probability density and
$\rho_2(x_{12})$ is a two-variable probability density.

To get a closed equation for $\rho_1(x_1)$, 
we use the approximation in which the two-variable probability
is written as the product of one-variable probabilities,
that is, we assume that the variables $x_1$ and $x_2$ are
statistically independent, $\rho_2(x_{12})=\rho_1(x_1)\rho_1(x_2)$.
This is a good approximation when the 
the particles are far from each other, as happens to a rarefied gas.
Equation (\ref{46}) becomes
\[
\frac{\partial}{\partial t}\rho_1(x_1) = 
\int w_{12}(x_{12} | x'_{12}) \times
\] 
\begin{equation}
\times \{\rho_1(x'_1)\rho_1(x'_2) - \rho_1(x_1)\rho_1(x_2) \} 
dx'_1 dx'_2 dx_2,
\label{47}
\end{equation}
which is the Boltzmann kinetic equation, except for the
transition rate which should be specified. This will
be done below. 

The production of entropy, given by (\ref{51}), becomes
\[
\Pi =  n\frac{k_{\mbox{\tiny B}}}2
\int \{\rho_1(x_1)\rho_1(x_2) - \rho_1(x'_1)\rho_1(x'_2)\}  \times 
\]
\begin{equation}
\times w_{12}(x'_{12} | x_{12})
\ln \frac{\rho_1(x_1)\rho_1(x_2)}{\rho_1(x'_1)\rho_1(x'_2)} dx_1dx_2dx'_1dx'_2.
\label{51a}
\end{equation}
Recalling that
\begin{equation}
\frac{dS}{dt} = \Pi,
\label{51b}
\end{equation}
we see that the right-hand side of equation (\ref{51a}) is identical
to the Boltzmann expression for $-dH/dt$.
Taking into account that $\Pi\geq0$, we conclude that $dS/dt\geq0$,
or that $dH/dt\leq0$, which is the Boltzmann H-theorem.

\section{Stochastic kinetic equation}

\subsection{Maxwell}

Here we provide a derivation of the Boltzmann
kinetic equation following the original lines of thought of
Maxwell and also that of Boltzmann but interpreting their reasoning as
stochastic. At the end, we set up a differential form of the
Boltzmann kinetic equation, obtained when the angles of deflection
of the molecules by collisions can be considered to be small.

The model used by Maxwell
to describe the kinetics of an ideal gas consisted of molecules
that move in straight lines in any possible direction
and speed, colliding with each other and with the walls of the container.
The velocities of the molecules are not arbitrary but are
in accordance with the well known velocity distribution 
introduced by Maxwell \cite{maxwell1860}
\begin{equation}
\rho({\bf v}) = b \,e^{-av^2/2}
\end{equation}
where $a$ and $b$ are constants, 
$v=|{\bf v}|$, and ${\bf v}$ is the velocity
${\bf v}$ of a molecule. A derivation
of the velocity distribution provided by Maxwell \cite{maxwell1867}
used reasonings of the stochastic type, and are as follows. 

We consider a gas consisting of several molecules of two types.
Let $\rho_2({\bf v}_1,{\bf v}_2)$ represent the final velocity 
probability density related to two types of molecules, A and B, 
where ${\bf v}_1$ and ${\bf v}_2$ denotes the velocities of A and B, respectively.
The velocities of the molecules are being
changed continuously due to the interaction between molecules,
such as collisions in the case of molecules modeled by rigid elastic bodies.
During a certain small interval of time $\tau$, let us suppose that the velocities
of two molecules A and B, change from $({\bf v}_1,{\bf v}_2)$ to
$({\bf v}_1^{\,\prime},{\bf v}_2^{\,\prime})$.
The probability of finding the velocities of A and B 
inside $d{\bf v}_1^{\,\prime} d{\bf v}_2^{\,\prime}$
around $({\bf v}_1^{\,\prime},{\bf v}_2^{\,\prime})$ at a certain instant of time
and inside $d{\bf v}_1 d{\bf v}_2$
around $({\bf v}_1,{\bf v}_2)$ at an earlier small interval of time $\tau$ is
\begin{equation}
\left\{w({\bf v}_1^{\,\prime},{\bf v}_2^{\,\prime} \,|\,{\bf v}_1,{\bf v}_2)
d{\bf v}_1^{\,\prime} d{\bf v}_2^{\,\prime} \,\tau \right\}
\rho_2({\bf v}_1,{\bf v}_2) d{\bf v}_1 d{\bf v}_2,
\label{8a}
\end{equation}
where the expression inside curly brackets
is the {\it conditional} probability of finding the velocities of A and B 
inside $d{\bf v}_1^{\,\prime} d{\bf v}_2^{\,\prime}$ around
$({\bf v}_1^{\,\prime},{\bf v}_2^{\,\prime})$ at a certain instant of time
$t$ {\it given} that they were $({\bf v}_1,{\bf v}_2)$ at an earlier
time $t-\tau$. The analogous expression for the reverse transition 
$({\bf v}_1^{\,\prime},{\bf v}_2^{\,\prime})\to({\bf v}_1,{\bf v}_2)$ is
\begin{equation}
\left\{w({\bf v}_1,{\bf v}_2\,|\,{\bf v}_1^{\,\prime},{\bf v}_2^{\,\prime})
d{\bf v}_1 d{\bf v}_2 \,\tau \right \}
\rho_2({\bf v}_1^{\,\prime},{\bf v}_2^{\,\prime})
d{\bf v}_1^{\,\prime}d{\bf v}_2^{\,\prime}.
\label{8b}
\end{equation}

If the gas is in a state of equilibrium, expression (\ref{8a})
should be equal to (\ref{8b}), that is,
\begin{equation}
w({\bf v}_1^{\,\prime},{\bf v}_2^{\,\prime} \,|\, {\bf v}_1,{\bf v}_2)
\rho_2({\bf v}_1,{\bf v}_2)
= w({\bf v}_1,{\bf v}_2\,|\,{\bf v}_1^{\,\prime},{\bf v}_2^{\,\prime})
\rho_2({\bf v}_1^{\,\prime},{\bf v}_2^{\,\prime}). 
\label{9}
\end{equation}
This is the well known detailed balance condition, or microscopic
reversibility, which is the condition of reversibility
of stochastic processes and identified as the condition
for thermodynamic equilibrium \cite{tome2015book}. 

The transition rate considered by Maxwell corresponds to
the collision of two molecules. Assuming that 
the kinetic energy before the collision equals that after
the collision, it follows that the transition rate
$w({\bf v}_1^{\,\prime},{\bf v}_2^{\,\prime}\,|\,{\bf v}_1,{\bf v}_2 )$
will be nonzero only when the conservation of energy is fulfilled,
that is,
\begin{equation}
m_1 v_1^{\prime\,2} + m_2 v_2^{\prime\,2} = m_1 v_1^2 + m_2 v_2^2,
\label{7}
\end{equation}
where $m_1$ and $m_2$ are the masses of molecules A and B, respectively.
At this point Maxwell assumes the transition rate and its reverse
to be equal, 
\begin{equation}
w({\bf v}_1^{\,\prime},{\bf v}_2^{\,\prime}\,|\,{\bf v}_1,{\bf v}_2)
=w({\bf v}_1,{\bf v}_2\,|\,{\bf v}_1^{\,\prime},{\bf v}_2^{\,\prime}).
\label{6}
\end{equation}
This result together with the equality (\ref{9})
allows us to write
\begin{equation}
\rho_2({\bf v}_1,{\bf v}_2) = \rho_2({\bf v}_1^{\,\prime},{\bf v}_2^{\,\prime}).
\label{8}
\end{equation}
Assuming that the velocities of two molecules are statistically
independent, it follows that
\begin{equation}
\rho_1({\bf v}_1)\rho_1({\bf v}_2)
= \rho_1({\bf v}_1^{\,\prime})\rho_1({\bf v}_2^{\,\prime}),
\label{8c}
\end{equation}
After taking the logarithm of both sides of equation (\ref{8c}),
\begin{equation}
\ln \rho_1({\bf v}_1) + \ln\rho_1({\bf v}_2)
= \ln \rho_1({\bf v}_1^{\,\prime}) + \ln\rho_1({\bf v}_2^{\,\prime}),
\label{8d}
\end{equation}
and comparing with the conservation of kinetic energy,
given by expression (\ref{7}), the
Maxwell distribution of velocities is obtained. 

\subsection{Boltzmann}

Boltzmann proposed the kinetic equation that bears his
name following a line of thought that can be
interpreted as stochastic reasonings,
as follows \cite{boltzmann1872}.
He first recognizes that the one-particle distribution
$\rho_1({\bf v}_1,t)$ changes in 
time due to the collisions between molecules, which change their
velocities. The difference $\rho_1({\bf v}_1,t+\tau)-\rho_1({\bf v}_1,t)$,
where $\tau$ is a small increment of time,
has a positive and a negative contribution.
The negative contribution comes from the
molecules that have velocities inside $d{\bf v}_1$
around ${\bf v}_1$ at time $t$ which change to velocities inside 
$d{\bf v}_2$ around ${\bf v}_2$ at $t+\tau$. 
This contribution is equal to expression (\ref{8a}), with the
understanding that $\rho_2({\bf v}_1,{\bf v}_2,t)$ now depends on time.
The positive contribution comes from the reverse situation and
is equal to expression (\ref{8b}).
By subtracting expressions (\ref{8b}) and (\ref{8a}),
and integrating over all possible
velocities ${\bf v}_2$, ${\bf v}_1^{\,\prime}$ and ${\bf v}_2^{\,\prime}$,
one obtains the difference $\rho_1({\bf v}_1,t+\tau)-\rho_1({\bf v}_1,t)$.
After dividing by $\tau$ and taking the limit $\tau\to0$, 
the following equation is obtained
\[
\frac{\partial}{\partial t}\rho_1({\bf v}_1) = \int
\{ w({\bf v}_1,{\bf v}_2\,|\,{\bf v}_1^{\,\prime},{\bf v}_2^{\,\prime})
\rho_2({\bf v}_1^{\,\prime},{\bf v}_2^{\,\prime})
\]
\begin{equation}
- w({\bf v}_1^{\,\prime},{\bf v}_2^{\,\prime}\,|\,{\bf v}_1,{\bf v}_2)
\rho_2({\bf v}_1,{\bf v}_2) \} d{\bf v}_2 d{\bf v}_1^{\,\prime} d{\bf v}_2^{\,\prime}.
\end{equation}

Like Maxwell, Boltzmann assumes that the probability transition rate
is nonzero when the kinetic energy, expressed by (\ref{7}) is conserved,
and that it is equal to its reverse, as given by equation (\ref{6}),
in which case the time evolution equation for $\rho_1({\bf v},t)$ becomes
\[
\frac{\partial}{\partial t}\rho_1({\bf v}_1) = \int
w({\bf v}_1,{\bf v}_2\,|\,{\bf v}_1^{\,\prime},{\bf v}_2^{\,\prime})\times
\]
\begin{equation}
\times
\{ \rho_2({\bf v}_1^{\,\prime},{\bf v}_2^{\,\prime})
-\rho_2({\bf v}_1,{\bf v}_2) \} d{\bf v}_2 d{\bf v}_1^{\,\prime} d{\bf v}_2^{\,\prime}.
\label{37}
\end{equation}
Assuming that ${\bf v}_1$ and ${\bf v}_2$ are statistically independent, 
equation (\ref{37}) becomes the Boltzmann kinetic equation \cite{boltzmann1872}
\[
\frac{\partial}{\partial t}\rho_1({\bf v}_1) = \int
w({\bf v}_1,{\bf v}_2\,|\,{\bf v}_1^{\,\prime},{\bf v}_2^{\,\prime})\times
\]
\begin{equation}
\times
\{ \rho_1({\bf v}_1^{\,\prime})\rho_1({\bf v}_2^{\,\prime})
- \rho_1({\bf v}_1)\rho_1({\bf v}_2) \}
d{\bf v}_2 d{\bf v}_1^{\,\prime} d{\bf v}_2^{\,\prime}.
\label{38}
\end{equation}
It should be remarked that in equilibrium the integrand vanishes 
and the Maxwell condition (\ref{8}) is recovered.

The reasonings employed above will also lead us to
the full master equation
\[
\frac{\partial \rho}{\partial t} = 
\frac1n \sum_{i<j} \int
w({\bf v}_i,{\bf v}_j\,|\,{\bf v}_i^{\,\prime},{\bf v}_j^{\,\prime})\times
\]
\begin{equation}
\times\{ \rho({\bf v}_i^{\,\prime},{\bf v}_j^{\,\prime})
- \rho({\bf v}_i,{\bf v}_j) \} d{\bf v}_i^{\,\prime} d{\bf v}_j^{\,\prime},
\label{39}
\end{equation}
where we have omitted, on the left-hand side, the dependence
of $\rho$ on all variables ${\bf v}_k$ and, on the right-hand side,
the dependence of $\rho$ on the variables ${\bf v}_k$, $k\neq i$ and
$k\neq j$. Equation (\ref{37}) follows from (\ref{39}) by integrating
over all velocities except ${\bf v}_1$.

\subsection{Transition rate}

The actual transition rate employed by Maxwell and Boltzmann,
in addition to the conservation of energy,
represented by relation (\ref{7}), also involved
the conservation of momentum
in a collision of two molecules, that is,
\begin{equation}
m_1 {\bf v}_1^{\,\prime} + m_2 {\bf v}_2^{\,\prime}
= m_1 {\bf v}_1 + m_2 {\bf v}_2.
\label{7a}
\end{equation}
From equation (\ref{7}) and (\ref{7a}) one obtains 
\begin{equation}
|{\bf v}_1^{\,\prime} - {\bf v}_2^{\,\prime}|
= |{\bf v}_1 - {\bf v}_2|.
\label{7b}
\end{equation}
Let us define the velocities ${\bf u}_{12}$ and ${\bf u}_{12}'$ 
of the center of mass before and after collision by
\begin{equation}
{\bf u}_{12} = r_2 {\bf v}_1 + r_1 {\bf v}_2,
\qquad
{\bf u}_{12}' = r_2 {\bf v}_1' + r_1 {\bf v}_2',
\end{equation}
where $r_1=m_2/(m_1+m_2)$ and $r_2=m_1/(m_1+m_2)$,
and the relative velocities ${\bf v}_{12}$ and ${\bf v}_{12}'$ before and
after the collision by
\begin{equation}
{\bf v}_{12} = {\bf v}_1 - {\bf v}_2,
\qquad
{\bf v}_{12}' = {\bf v}_1' - {\bf v}_2'.
\end{equation}
The two conservation laws (\ref{7a}) and (\ref{7b})
are equivalent to
\begin{equation}
{\bf u}_{12}'={\bf u}_{12},
\end{equation}
and $v_{12}'=v_{12}$, respectively. This last relation can be written in the form
\begin{equation}
{\bf v}'_{12}=v_{12}{\bf e},
\end{equation}
where ${\bf e}$ is a unit vector parameter. 
Denoting by $\alpha$ the angle between ${\bf e}$ and
the unit vector ${\bf v}_{12}$, we may write
\begin{equation}
{\bf e} = {\bf z}\cos\alpha + \mbox{\boldmath$\theta$}\sin\alpha\cos\beta
+ \mbox{\boldmath$\phi$}\sin\alpha\sin\beta,
\end{equation}
where the unit vectors {\boldmath$\theta$} and {\boldmath$\phi$} together
with ${\bf z}={\bf v}_{12}/v_{12}$ form a spherical orthogonal basis.

Regarding the vector ${\bf e}$ as a  random unit vector with a given
probability ${\cal P}({\bf e})d{\bf e}={\cal P}(\alpha,\beta)d\alpha d\beta$,
the rate of the transition
$({\bf v}_1,{\bf v}_2)\to({\bf v}_1',{\bf v}_2')$ is 
\begin{equation}
w({\bf v}_1',{\bf v}_2'|{\bf v}_1,{\bf v}_2) = a v_{12}\! \int\! 
\delta ({\bf u}_{12}' - {\bf u}_{12})
\delta ({\bf v}_{12}' - v_{12}{\bf e}) {\cal P}({\bf e})d{\bf e}.
\label{71}
\end{equation}
To fulfill the condition that the transition rate is equal to its
reverse, given by (\ref{6}), ${\cal P}({\bf e})$ is chosen
to be independent of the azimuthal angle $\beta$.

Following Maxwell \cite{maxwell1867}, we are choosing the pre-factor as
proportional to $v_{12}$. The factor $a$ is a constant having the
unit of frequency that sets the
scale of time. Replacing expression (\ref{71}) in 
the right-hand side of (\ref{38}), this term becomes 
\[
a v_{12} \int \{
  \rho_2({\bf v}_1+r_1(v_{12}{\bf e}-{\bf v}_{12}),
{\bf v}_2-r_2(v_{12}{\bf e}-{\bf v}_{12}) )
\]
\begin{equation}
- \rho_2({\bf v}_1,{\bf v}_2) \}{\cal P}({\bf e})d{\bf e}.
\label{61}
\end{equation}

The full master equation is
\begin{equation}
\frac{\partial \rho}{\partial t} = \frac{a}n \sum_{i<j} \int v_{ij} 
\{ \rho({\bf v}_i^{\,\prime},{\bf v}_j^{\,\prime})
- \rho({\bf v}_i,{\bf v}_j) \} {\cal P}({\bf e}_{ij})d{\bf e}_{ij},
\label{39a}
\end{equation}
where 
\begin{equation}
{\bf v}_i^{\,\prime} = {\bf v}_i+r_1(v_{ij}{\bf e}_{ij}-{\bf v}_{ij}),
\end{equation}
\begin{equation}
{\bf v}_j^{\,\prime} = {\bf v}_j-r_2(v_{ij}{\bf e}_{ij}-{\bf v}_{ij}),
\end{equation}
and we have omitted, on the left-hand side, the dependence
of $\rho$ on all variables ${\bf v}_k$ and, on the right-hand side,
the dependence of $\rho$ on the variables ${\bf v}_k$, $k\neq i$ and
$k\neq j$. 

\subsection{Differential form}

The angle between the relative velocity of two molecules 
before the collision, ${\bf v}_{12}$, and after the collision,
${\bf v}'_{12}=v_{12}{\bf e}$ is understood as the deflection
angle. For hard spheres the deflection angle is expected to be
equally probable. For other types of collision, such as that between
soft spheres, one expects a small angle to be more probable.
That is, we expect the deviation $\Delta {\bf v}_{12}=v_{12}{\bf e}-{\bf v}_{12}$
to be small, in general, and we may
expand the integrand of (\ref{61}) in powers of the deviation.
Up to second order, the expansion reads 
\begin{equation}
a \int\left(\Delta {\bf v}_{12} \cdot {\bf D}_{12}\rho + \frac1{v_{12}} 
(\Delta{\bf v}_{12} \cdot {\bf D}_{12})^2 \rho\right)
{\cal P}({\bf e})d{\bf e},
\label{59}
\end{equation}
where
\begin{equation}
{\bf D}_{12}\rho = v_{12}
\left(\frac{\partial\rho}{\partial{\bf v}_{12}}\right)_{v_{12}},
\end{equation}
and the derivations are performed in such a way that
$v_{12}$ is kept constant, or, in an equivalent form,
\begin{equation}
{\bf D}_{12}\rho = v_{12}
\left(-r_1\frac{\partial\rho}{\partial{\bf v_1}}
+ r_2\frac{\partial\rho}{\partial{\bf v}_2}\right)_{v_{12},{\bf u}_{12}},
\end{equation}
and the derivations are performed with
$v_{12}$ and ${\bf u}_{12}$ kept constant.

Taking into account that ${\bf v}_{12}$ is orthogonal to  
${\bf D}_{12}\rho$ it follows that 
$\Delta{\bf v}_{12} \cdot {\bf D}_{12}\rho=
v_{12}{\bf e} \cdot {\bf D}_{12}\rho$
and expression (\ref{59}) becomes
\begin{equation}
a v_{12} \int\left({\bf e} \cdot {\bf D}_{12}\rho + \frac12 
({\bf e} \cdot {\bf D}_{12})^2\rho\right)
{\cal P}({\bf e})d{\bf e}.
\end{equation}
Using the property that ${\cal P}({\bf e})$ does not depend on the
azimuthal angle $\beta$ and that ${\bf D}_{12}\rho$ is orthogonal to ${\bf z}$,
the first integral vanishes and we are left with
\begin{equation}
\frac12 a v_{12} \int \left(({\bf e} \cdot {\bf D}_{12})^2\rho\right)
{\cal P}({\bf e})d{\bf e}.
\label{59a}
\end{equation}

Let us write ${\bf D}_{12}$ in spherical coordinates
\begin{equation}
{\bf D}_{12} \rho = 
\mbox{\boldmath$\theta$}\frac{\partial\rho}{\partial\theta}
+ \frac{\mbox{\boldmath$\phi$}}{\sin\theta}\frac{\partial\rho}{\partial\phi},
\end{equation}
Replacing these expressions in (\ref{59a}), and performing the integral, it becomes
\begin{equation}
\frac12 b a v_{12} {\bf D}_{12} \cdot {\bf D}_{12}\rho 
= \frac12 b a v_{12} D_{12}^2\rho,
\label{59b}
\end{equation}
where 
\begin{equation}
b=\int (\sin\alpha\cos\beta)^2{\cal P}(\alpha)d\alpha d\beta.
\end{equation}
The explicit form of $D^2_{12}\rho$ is
\begin{equation}
D_{12}^2 \rho = \frac{1}{\sin\theta}\left(
\frac{\partial}{\partial\theta} \sin\theta\frac{\partial\rho}{\partial\theta}\right)
+ \frac{1}{\sin^2\theta} \frac{\partial^2\rho}{\partial\phi^2}.
\end{equation}

The master equation (\ref{39}) acquires the differential form 
\begin{equation}
\frac{\partial \rho}{\partial t} = \frac{ba}{2n} \sum_{i<j} 
 v_{ij} D_{ij}^2\rho,
\label{40}
\end{equation}
which is an equation of the Fokker-Planck type. 

The rate of entropy production $\Pi$ can be determined
from $\Pi=dS/dt$ because, in the present case, the entropy
flux $\Phi$ vanishes identically. From the definition of
entropy 
\begin{equation}
S = - k_{\mbox{\tiny B}}\int \rho \ln\rho \, d{\bf v},
\end{equation}
where $d{\bf v}=d{\bf v}_1 \ldots d{\bf v}_n$ one finds
\begin{equation}
\frac{dS}{dt} = - k_{\mbox{\tiny B}}\int \frac{\partial\rho}{\partial t}
\ln\rho \, d{\bf v}.
\label{41}
\end{equation}
After substituting (\ref{40}) in (\ref{41}) and performing
an integration by parts, one reaches the result for the
rate of entropy production
\begin{equation}
\Pi = nk_{\mbox{\tiny B}}
\frac{ba}{2}  \int \frac{v_{ij}}{\rho}({\bf D}_{12}\rho)^2 \, d{\bf v},
\end{equation}
which, clearly, is a nonnegative quantity.

\section{Liouville equation}

\subsection{Transition rate}

The Liouville equation gives the evolution of the
probability density $\rho(x,t)$ in phase space 
of a system that evolves in time in a deterministic way.  
Let us consider a system with $n$ degrees of freedom described
by a Hamiltonian ${\cal H}(x)$, defined in the phase space $x=(q,p)$,
where $q$ and $p$ denote vectors with components $q_i$
$p_i$, respectively.
The evolution of a point in phase space is given by
\begin{equation}
\frac{dq}{dt} = \frac{ \partial{\cal H}}{\partial p},
\qquad\qquad
\frac{dp}{dt} = -\frac{\partial{\cal H}}{\partial q},
\label{53}
\end{equation} 
where the partial derivatives denote vectors with components 
$\partial{\cal H}/\partial q_i$ and $\partial{\cal H}/\partial p_i$.
The Liouville equation is
\begin{equation}
\frac{\partial \rho}{\partial t} = \{{\cal H},\rho\},
\label{55}
\end{equation}
where $\{A,B\}$ are the Poisson brackets.

Usually, the Liouville equation is introduced by the 
use of the Liouville theorem according to which 
the volume of a certain region in phase space is preserved
under the Hamiltonian dynamics (\ref{53}). If $x$ represents the system
at time $t$ and $x'$ at time $t'$, the Liouville
theorem is expressed by $dx=dx'$. 
In addition, one postulates that the probability of 
two regions that are mapped onto each other by the
Hamiltonian dynamics are equal,
that is, $\rho(x,t)dx=\rho(x',t')dx'$. This postulate 
together with Liouville theorem gives $\rho(x,t)=\rho(x',t')$,
from which follows the Liouville equation (\ref{55}).

Here, instead, we derive the Liouville equation from equation (\ref{20})
by the use of a transition rate $w(x'|x)$ which is nonzero only when 
$x=(q,p)$ evolves to a single state $x'=(q',p')$.
For a small interval of time $\tau$, the new
position $(q',p')$ in phase space is
\begin{equation}
q' = q + \frac{\partial{\cal H}}{\partial p}\tau,
\qquad\qquad
p' = p - \frac{\partial{\cal H}}{\partial q}\tau.
\end{equation}
In this case, the transition rate is given by
\begin{equation}
w(x'|x) \tau = 
\delta(q' - q - \frac{\partial{\cal H}}{\partial p}\tau)\,
\delta(p' - p + \frac{\partial{\cal H}}{\partial q}\tau),
\end{equation}
which replaced in the right-hand side of equation (\ref{20}) gives
\begin{equation}
\frac{1}{\tau} \{\rho(q,p) -
\rho(q + \frac{\partial{\cal H}}{\partial p}\tau,
p - \frac{\partial{\cal H}}{\partial q}\tau) \}.
\label{52}
\end{equation}
Taking the limit $\tau\to0$, we reach the result
\begin{equation}
\frac{\partial\rho}{\partial t} =
- \left(
 \frac{\partial\rho}{\partial q} \cdot \frac{\partial{\cal H}}{\partial p}
-\frac{\partial\rho}{\partial p} \cdot \frac{\partial{\cal H}}{\partial q}
\right),
\label{54}
\end{equation}
which is the Liouville equation (\ref{55}).

The time derivative of the entropy $S$, given by (\ref{18}), 
vanishes identically and we conclude that $S$ is constant in time.
To see this, it suffices to replace the Liouville equation
in the expression for $dS/dt$ and perform an integration by parts.
Considering that the Liouville equation is understood as 
describing an isolated system, this result does not
envisage the increase in entropy as required by thermodynamics
of isolated systems. In contrast, the master equation that
we have set up for isolated systems can predict an increase in entropy.

\subsection{Jeans equation}

Let us consider an approximation to the Liouville equation
similar to that we have employed when we considered the
variables corresponding to two particles as statistically
independent.
To this end, we start by writing down the evolution equation for
the one particle probability density $\rho_1(x_1)$.
Considering a Hamiltonian of the type
\begin{equation}
{\cal H} = \sum_i \frac{p_i^2}{2m} + \sum_{(ij)}{\cal H}_{ij}(q_i,q_j),
\label{56}
\end{equation}
and integrating the Liouville equation over all variables except $x_1$,
we find
\begin{equation}
\frac{\partial f_1}{\partial t} =
- \frac{\partial f_1}{\partial q_1} \frac{p_1}{m}
+ \int \frac{\partial f_{12}}{\partial p_1}
\frac{\partial{\cal H}_{12}}{\partial q_1} dq_2dp_2.
\label{57}
\end{equation}
where $f_1=n\rho_1$ and $f_{12}=n(n-1)\rho_{12}$, and
$n$ is the number of degrees of freedom.  

Assuming that the variables $x_1$ and $x_2$
are statistically independent, that is,
using the approximation $\rho_2(x_1,x_2)=\rho_1(x_1)\rho_1(x_2)$,
which yields $f_2(x_1,x_2)=f_1(x_1)f_1(x_2)$ for large $n$, we get
\begin{equation}
\frac{\partial f_1}{\partial t} =
- \frac{\partial f_1}{\partial q_1} \frac{p_1}{m}
+\frac{\partial f_1}{\partial p_1}  
\frac{\partial\psi_1}{\partial q_1},
\label{58}
\end{equation}
where 
\begin{equation}
\psi_1 =  \int f_1(q_2,p_2) {\cal H}_{12} dq_2 dp_2.
\label{58a}
\end{equation}
Equation (\ref{58}) was proposed by Jeans \cite{jeans1915,jeans1919}
in the context of stellar dynamics in which case the ${\cal H}_{ij}$
is the gravitational potential between particles $i$ and $j$.
It is also known as the collisionless Boltzmann equation and sometimes
the Vlasov equation \cite{henon1982}.

To determine the production of entropy predicted by Jeans equation,
we may use the Boltzmann entropy 
\begin{equation}
S = - k_{\mbox{\tiny B}} \int f_1 \ln f_1 dq_1dp_1,
\label{18a}
\end{equation}
which comes from equation (\ref{18}) by assuming the same
approximation that we have used to derive the Jeans equation.
Again, the time derivative of the entropy $S$ 
related to the Jeans equation vanishes identically and 
$S$ is constant in time. To see this, it suffices to replace the Jeans equation
in the expression for $dS/dt$ and perform an integration by parts.
Therefore, not only the Liouville equation but also the
approximate Jeans equation predicts an invariance of entropy
in time.

\section{Conclusion}

We have interpreted the dynamics of the kinetic 
theory of gases of Maxwell and Boltzmann
as a stochastic dynamics.
Accordingly, a trajectory emerging from a given point
of the phase space is not unique, as happens with a deterministic
dynamics obeying the laws of classical motion, 
but may be split into 
several trajectories, each one with a given probability
of occurrence. When the dynamics is deterministic, it
leads to the Liouville equation, which gives no
generation of entropy. On the other hand, if the dynamics
is stochastic, as with the Boltzmann equation, there is a production 
of entropy. 
Considering that, according to thermodynamics,
the entropy increases in isolated systems, the appropriate
description of these systems could be given by an equation
which is a combination of the right hand-side
of Liouville equation (\ref{55})
and the right-hand side of the Boltzmann kinetic equation (\ref{20i}),
\begin{equation}
\frac{\partial\rho}{\partial t} = \{{\cal H},\rho\} 
+ \int w(x | x') \{\rho(x') - \rho(x) \}dx'.
\label{62}
\end{equation}

The irreversible character of the second law of thermodynamics,
expressed by the increase of entropy of isolated systems $dS/dt\geq0$,
comes from the Boltzmann part of equation (\ref{62}).
Indeed, the production of entropy $\Pi$ associated with the evolution of
$\rho$ given by equation (\ref{62}) is expressed by equation (\ref{51}),
which is nonnegative implying $dS/dt=\Pi\geq0$. This is 
equivalent to the Boltzmann H-theorem.

It is worth asking what types of interacting forces 
should be included in the Liouville part of equation (\ref{62}),
producing no entropy, and what types should be
in the Boltzmann part, giving rise
to entropy production in a nonequilibrium regime.
A possible answer is to be found in the
Lyapunov exponent, which says whether the trajectories
determined by a force will diverge exponentially or not when
they originate from nearby states.
If they diverge, as happens to the collision of hard spheres, 
in which case the Lyapunov exponent is positive,
this type of force should be included in the Boltzmann part;
otherwise, in the Liouville part.

The stochastic character of the dynamics is embodied 
in $P({\xi})$, introduced as an {\it a priori} probability distribution, 
from which one obtains the transition rate. In the case of
the Boltzmann equation (\ref{38}), this probability distribution,
called ${\cal P}({\bf e})={\cal P}(\alpha,\beta)$, is identified
as being proportional to the collision differential cross section between 
two colliding molecules, and the polar angle $\alpha$ is identified as
the deflection angle. 

We have also shown that the Boltzmann equation can be reduced to
a differential form, in which case it can be understood
as an equation of the Fokker-Planck type. This reduction
is obtained when the deviation between 
relative velocities of two molecules
before and after a collision is small. The magnitude
of this deviation is given by the scattering angle $\alpha$,
which should then be small.  The differential operator
${\bf D}_{ij}$ changes the velocities of two molecules but
preserves the sum of their kinetic energies and the sum of their momenta. 
Therefore, the total kinetic energy and the total momentum
of the molecules is preserved by the differential
form of the Boltzmann equation. 

The aim of the kinetic theory of the nineteenth century
was the derivation
of the macroscopic properties of gases from the microscopic
laws that govern the motion of molecules. It was tacitly understood
that the derivation should be founded on the use of deterministic
laws of mechanics, only, that is, a purely mechanical derivation.
However, this aim was not fully accomplished.
The purely mechanical derivation was replaced by a derivation 
containing, sometimes implicitly, probabilistic and stochastic reasonings.
The crucial step in our stochastic approach 
was the introduction of the {\it a priori}
probability distribution ${\cal P}(\xi)$ of the 
variable $\xi$ which describes the stochastic motion.
If this probability distribution could be provided by
the laws of mechanics, that is, from pure mechanics,
then the original aim of kinetic theory would be fully accomplished.

The reasonings of Maxwell and Boltzmann tacitly employed
two important assumptions related to irreversibility
that were later made explicit by Boltzmann
himself, by Jeans, and by the Ehrenfests.
One of them was the assumption of statistical independence between the 
dynamic variables of one molecule and those of another molecule.
We employed this assumption, which is in fact an approximation,
to get the Boltzmann equation (\ref{38}) from (\ref{37}), or from
the full master equation  (\ref{39}), and to get the Jeans equation
(\ref{58a}) from the Liouville equation (\ref{55}).
The Ehrenfests \cite{ehrenfest1912} called this assumption
the hypothesis of molecular disorder ({\it molekularen Unordnung}),
but this cannot be seen as the source of irreversibility and production of entropy.
When this assumption is applied to the Liouville equation, the resulting 
Jeans equation produces no entropy.

The second assumption concerns the transition rate.
The frequency of collisions of two molecules which changes their
velocities is assumed to be independent of the position of the molecules. 
In addition, the frequency of collisions is the same if
the velocities of the molecules are interchanged. 
This assumption introduces probabilistic
and stochastic elements into the approach and is the source 
of irreversibility and entropy production. 
This assumption allows us to write the rate in terms
of the velocities only, and it gives the property (\ref{6}). 
This crucial assumption was
called the collision number hypothesis ({\it Stosszahlansatz})
by the Ehrenfests \cite{ehrenfest1912}.

Boltzmann \cite{boltzmann1896} stated that the two assumptions above were 
consequences of the state of the gas being molecularly disordered
({\it molekular Ungeordnet}) \cite{boltzmann1896}.
Jeans \cite{jeans1904} distinguished the two assumptions
but considered them as consequences of the assumption of molecular chaos,
and the Ehrenfests \cite{ehrenfest1912} not only distinguished the
two assumptions but called them by different names.


\end{document}